\documentclass[twocolumn]{jpsj3}
\usepackage{txfonts}
\usepackage{times}
\usepackage{color}

\title{Antiferromagnetism in Iron-Based Superconductors: Selection of Magnetic Order and Quasiparticle Interference}

\author{Ilya Eremin$^{1}$, Johannes Knolle$^{2}$, Rafael M. Fernandes$^{3}$, J\"org Schmalian$^{4}$, and Andrey V. Chubukov$^{5}$
}
\inst{$1$-Institut f\"ur Theoretische Physik III, Ruhr-Universit\"at Bochum, D-44801 Bochum, Germany \\
$2-$Max Planck Institute for the Physics of Complex Systems, D-01187 Dresden, Germany \\
$3-$School of Physics and Astronomy, University of Minnesota, Minneapolis, 55455, USA \\
$4-$Institut f\"ur Theorie der Kondensierten Materie, Karlsruher Institut f\"ur Technologie, DE-76131 Karlsruhe, Germany
$5-$Department of Physics, University of Wisconsin-Madison, Madison, Wisconsin 53706, USA \\
}
\abst{
The recent discovery of superconductivity in the iron-based layered pnictides with T$_c$ ranging between 26 and 56K  generated enormous interest  in the physics of these materials. Here, we review some of the peculiarities of the antiferromagnetic order in 
the iron pnictides, including the selection of the stripe magnetic order and the formation of the 
Ising-nematic state in the unfolded BZ within 
an itinerant description. In addition we analyze the properties of the quasiparticle interference 
spectrum in the parent antiferromagnetic phase.}

\kword{iron-based superconductors, magnetic fluctuations, quasiparticle interference}

\begin{document}
\maketitle

\section{Introduction}

The comprehensive  understanding of the relationship between magnetism and
superconductivity in the Fe-based superconductors, discovered in 2008 by Hideo Hosono and collaborators\cite{kamihara2008}, ultimately requires an analysis of the magnetic ground states in these compounds and their evolution with doping. In particular, the origin of magnetism in the FeSC parent compounds is hotly debated 
since it is believed that the same magnetic interactions that drive the magnetic ordering also produce the Cooper-pairing\cite{mazin}. The phase diagram of ferropnictides (FPs) is similar to
high-T$_c$ cuprates 
as it contains an antiferromagnetic (AF) phase in close proximity to the superconducting (SC) one. Most 
FPs exhibit an AF state at low carrier
concentrations,
whose suppression with doping, pressure, or disorder allows for the emergence of
superconductivity. This shows strong similarities to the generic cuprate phase diagram
and is evidence for the interplay 
between magnetism and superconductivity in the Fe-based
materials. There are two important distinctions,
however. First, parent compounds of 
FPs are antiferromagnetic
{\it metals}, and second, the superconducting pairing symmetry in most of the materials is, most likely, an extended $s$-wave,
with or without nodes.\cite{mazin}
The electronic structure of 
the parent FPs in the normal state has
been measured by angle-resolved photoemission (ARPES)\cite{Kaminski2008,Terashima2009,Borisenko2009,Feng2009,Shen2008,Ding2011}
and by 
quantum oscillations\cite{Coldea2008,Carrington}. Both agree largely with {\it ab-initio} band structure calculations\cite{Singh2008,Boeri2008}. It consists of two
quasi-two-dimensional near-circular hole pockets
of unequal size, centered around the $\Gamma$-point (0,0), and two
quasi-2D 
elliptical electron pockets centered around (0,$\pi$)
and $(\pi,0)$ points in the unfolded Brillouin zone (BZ)
which includes only Fe atoms. 
Due to the tetragonal symmetry, the
two electron pockets transform into each other under rotation
by 90$^o$. In the folded BZ, which is used for experimental
measurements because of 
the two nonequivalent As positions
with respect to 
the Fe plane, both electron pockets are centered
around $(\pi,\pi)$. The dispersions near electron pockets
and near hole pockets are reasonably close to each other
apart from the sign change, {\it i.e.}, there is a substantial degree
of nesting between hole and electron bands. One has to mention that nesting of electron and hole bands is not always present in iron-based superconductors and we comment on these systems at the end of this chapter.

Here we 
review theoretically the formation of antiferromagnetic order in parent FPs and its consequences for the 
electronic structure, as well as for the appearance of Ising-nematic order above the magnetic transition.  Some of the results appeared previously in Refs.\cite{Fernandes2011,Eremin2010} In addition, we address the quasiparticle interference 
spectra in the magnetically ordered state and compare them to the experimental data. We will only focus on 
metallic FeAs materials, for which the weak-coupling analysis seems to be applicable. Neutron
scattering measurements on parent FeAs pnictides have
revealed the ordering momentum in the unfolded BZ to be
either $(0,\pi)$, or $(\pi,0)$, i.e. 
the magnetic order consists of ferromagnetic
chains along one crystallographic direction 
and antiferromagnetic chains along the other direction.
Such magnetic order emerges in the $J_1-J_2$ model of localized
spins with 
exchange interactions between nearest and next-nearest
neighbors $J_1$ and $J_2$, respectively, for $J_2>0.5J_1$\cite{Chandra1990,Si2008,Xu2008,Yildirim2008,Uhrig2009}. However, here we discuss an alternative scenario which assumes that parent FPs are
good metals made of itinerant electrons, and antiferromagnetic
order is of 
spin-density wave (SDW) type. 
Indeed, optical conductivity measurements observe the transfer of spectral weight from the Drude peak to a mid-infrared peak, consistent with itinerant electrons giving rise to AF order \cite{Uchida2009}.
The nesting-driven mechanism is known to give rise to the incommensurate
AF in Cr \cite{Rice1970,Keldysh1965}.  Given the electronic structure of
FPs, it is natural to assume that AF order emerges, at least
partly, due to near nesting between the dispersions of holes
and electrons. This is confirmed by ab-initio analysis of the total energy in the antiferromagnetic state
which shows that the main energy gain with respect to the paramagnetic state comes from regions of the BZ where electron and hole pockets reside \cite{Andersen2011}. 
Furthermore, angle-resolved photoemission spectroscopy (ARPES) find a direct relationship between nesting and the onset of AF \cite{Liu2010}. An incommensurate AF order is also observed by neutron diffraction for some doping values \cite{Pratt2011}.
Several groups 
have explored different aspects of this itinerant SDW model over the last years.\cite{cvetkovic2009,Chubukov2008,Brydon2009,wang2009,platt2009,Fernandes2011}

%
%

\section{Magnetic order in ferropnictides}\label{sec1.2}

\subsection{Magnetic frustration}

As pointed out in the introduction the magnetic order in the 
FPs  was originally detected by neutron scattering\cite{Cruz2008} and
$\mu$sR experiments\cite{Klauss2008}. The magnetic transition temperature varies slightly from compound to compound and is of the order of
$T_{N} \sim 150$K. In real space the magnetic ordering consists of ferromagnetic chains along one crystallographic direction in the Fe square lattice which are coupled antiferromagnetically. In  momentum space the order can be characterized by 
the wave-vectors ${\bf Q}_1 = (\pi,0)$ or ${\bf Q}_2 = (0,\pi)$. Within the localized scenario this order emerges in the context of the $J_1-J_2$ model \cite{Chandra1990} for $J_2> 0.5 J_1$ once quantum fluctuations are taken into account. In the following we review how this order appears in the itinerant picture.

\begin{figure}[h]
\begin{center}
\includegraphics[width=7cm]{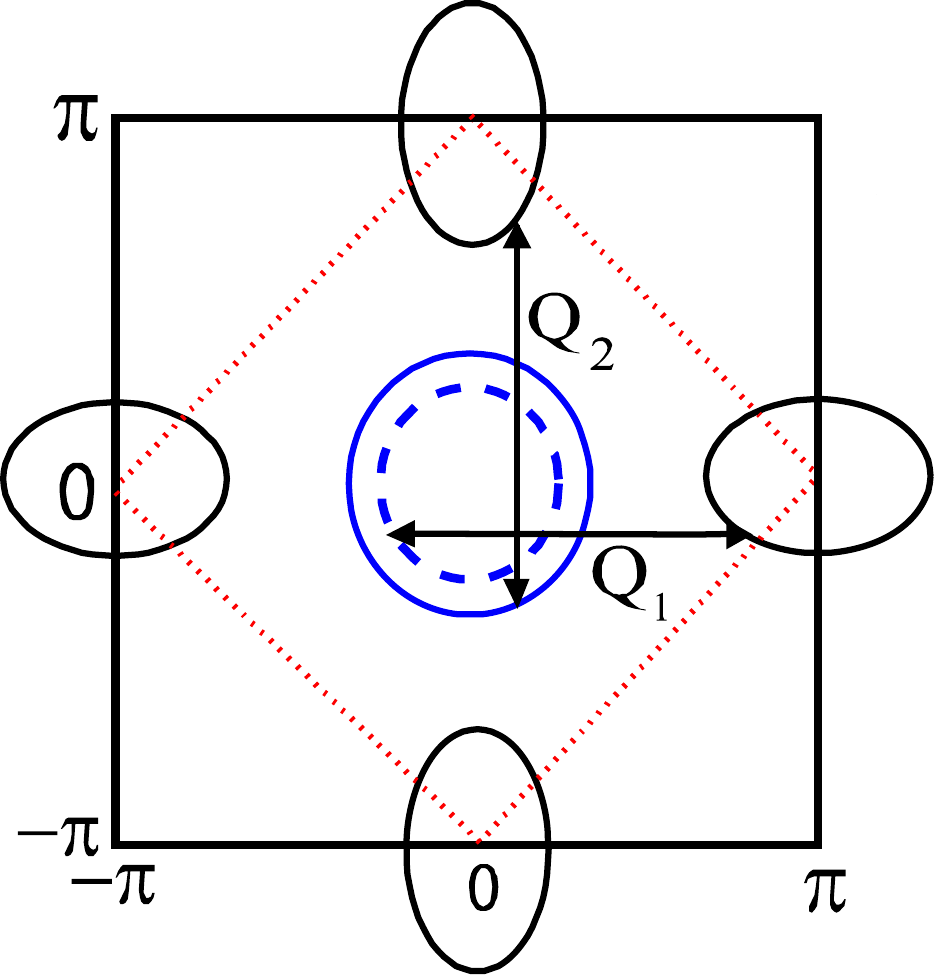}
\end{center}
\caption{(Color online) Schematic Fermi surface of ferropnictides in the unfolded Fe-based Brillouin zone with two circular hole-like pockets centered around the $\Gamma$-point and two elliptical electron-like pockets located around the $(\pi,0)$ and $(0,\pi)$ point of the BZ, respectively. ${\bf Q_1}$ and ${\bf Q_2}$ represent two nesting wave vectors.}
\label{FS}
\end{figure}
The schematic Fermi surface (FS) in the normal state of FPs is reproduced in Fig.\ref{FS} for the unfolded 
 BZ (i.e the square-lattice BZ). As we pointed out in the introduction the elliptical electron bands and nearly circular hole bands are nearly nested.
For the Fermi surface topology of FPs it means that there are two nesting wave
vectors ${\bf Q_1}=(\pi,0)$ and ${\bf Q_2}=(0,\pi)$ between the hole- and electron-like pockets. For the idealized case of zero ellipticity of the electron pockets and equal masses for the electron and hole bands the situation is similar to the half-filled Hubbard model with nearest neighbor hopping. In particular, the susceptibility in the particle-hole channel diverges logarithmically  as it usually does in the particle-particle Cooper-channel. A renormalization group analysis shows that the leading instability is magnetic.\cite{cvetkovic2009,Chubukov2008} However, here it occurs at two wave vectors ${\bf Q_1}$ and ${\bf Q_2}$,
leaving open the question of how only one of the two ordering vectors is selected, as it is experimentally observed.

\begin{figure}[h]
\begin{center}
\includegraphics[width=9cm]{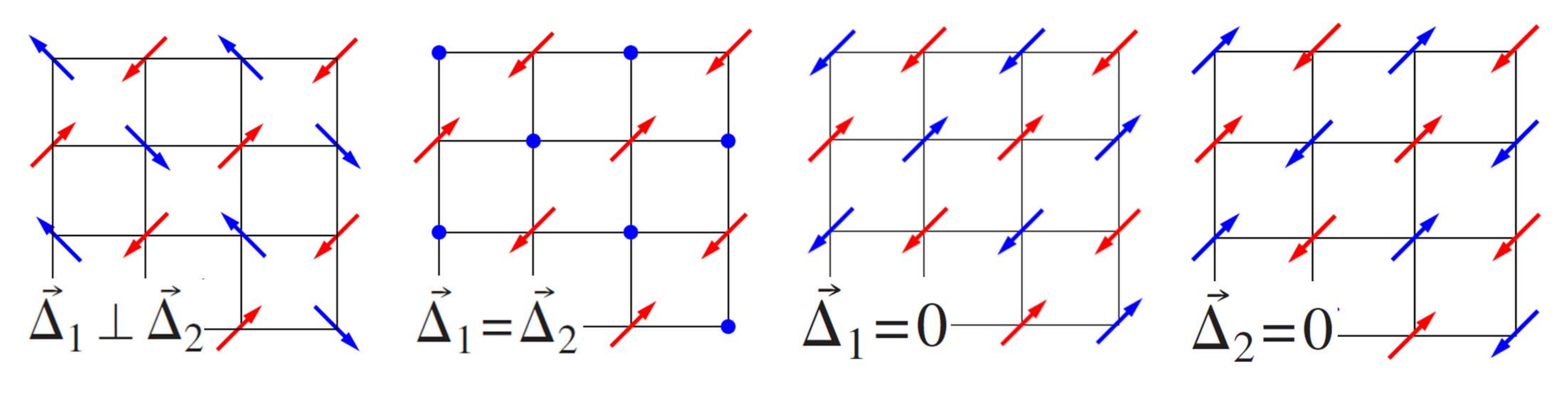}
\end{center}
\caption{(Color online) Possible real space orderings of the magnetic state. Either one of the last two corresponds to the experimentally realized one \cite{Eremin2010}.}
\label{MagOrdRealSpace}
\end{figure}

To formulate the problem in a formal way let us start with a generic spin configuration described by two mean-field SDW order parameters $\overrightarrow{\Delta}_i$ for each of the wave vector ${\bf Q_i}$ in the form
\begin{equation}
\overrightarrow{S} ({\bf R}) = \overrightarrow{\Delta}_1 e^{i {\bf Q_1 R}} + \overrightarrow{\Delta}_2 e^{i {\bf Q_2 R}}.
\label{SpinOperatorTwoOP}
\end{equation}
For such a configuration the Fe lattice decouples into two 
interpenetrating antiferromagnetically ordered sublattices with magnetizations $\overrightarrow{\Delta}_1+\overrightarrow{\Delta}_2$ and $\overrightarrow{\Delta}_1-\overrightarrow{\Delta}_2$.  However, neither the angle between the two 
Neel vectors is fixed nor their magnitudes. For example, in Fig.\ref{MagOrdRealSpace} we show four possible orderings for a generic
$\overrightarrow{S}({\bf R})$ out of many possibilities. The last two configurations with one of the $\overrightarrow{\Delta}_i$ 
vanishing refer to the experimentally observed ones.

Without loss of generality one can assume that one of the hole pockets 
interacts stronger with the two electron-like pockets than the other hole pocket. Therefore it is useful to consider a model consisting of a single circular hole FS centered around the  $\Gamma$-point ($\alpha$-band)
and two elliptical electron Fermi surface pockets centered around 
the $(\pm \pi, 0)$ and $(0, \pm \pi)$ points in the unfolded BZ ($\beta$-bands):
\begin{eqnarray}
\label{eqH}
\lefteqn{H_2  =} && \nonumber\\
&& \sum_{\mathbf{p}, \sigma} \left[ \varepsilon^{\alpha_1}_{\mathbf{p}} \alpha_{1\mathbf{p}  \sigma}^\dag \alpha_{1\mathbf{p} \sigma}
+ \varepsilon^{\beta_{1}}_{\mathbf{p}} \beta_{1\mathbf{p} \sigma}^\dag \beta_{1\mathbf{p} \sigma} + \varepsilon^{\beta_{2}}_{\mathbf{p}} \beta_{2 \mathbf{p} \sigma}^\dag \beta_{2\mathbf{p} \sigma} \right]. \nonumber\\
\end{eqnarray}
Here, $\varepsilon^{\alpha_{1}}_{\mathbf{p}} = - \frac{\hbar^2 p^2}{2 m_1}
+\mu $ and  $\varepsilon^{\beta_1}_{\mathbf{p}}= \frac{\hbar^2p_x^2 }{2 m_{x}} + \frac{\hbar^2 p_y^2 }{2 m_{y}} -\mu $, $\varepsilon^{\beta_2}_{\mathbf{p}}= \frac{\hbar^2 p_x^2}{2 m_{y}} +\frac{\hbar^2 p_y^2}{2 m_{x}} -\mu $ are the dispersions of 
the hole and electron bands.
The momenta of $\alpha-$ fermions are counted from $(0,0)$, 
whereas the momenta of the
$\beta_1-$ and $\beta_2-$fermions are counted  from $(0,\pi)$ and ($\pi,0$).

One then projects all possible electronic interactions in the SDW channel. 
According to the terminology of Ref.\cite{Chubukov2008}, the dominant contributions are the density-density interactions
between $\alpha$ and $\beta$ fermions in the form
\begin{eqnarray}
&&H_{4}=U_1 \sum  {\alpha}^{\dagger}_{1{\bf p}_3 \sigma}
{\beta}^{\dagger}_{j{\bf p}_4 \sigma'}  {\beta}_{j{\bf p}_2 \sigma'} {\alpha}_{1{\bf p}_1
\sigma}  + \nonumber \\
&& \frac{U_3}{2}~ \sum
\left[{\beta}^{\dagger}_{j{\bf p}_3 \sigma} {\beta}^{\dagger}_{j{\bf p}_4 \sigma'}
{\alpha}_{1{\bf p}_2 \sigma'} {\alpha}_{1{\bf p}_1 \sigma} + h.c \right].
 \label{eq:2}
\end{eqnarray}
At this level we neglect the potential angular dependencies of $U_1$ and $U_3$ along the FSs which arise due to 
orbital contents of the the Fermi pockets,
which we will discuss later.

The two SDW order parameters are expressed as
$\overrightarrow{\Delta}_1 \propto \sum_{\bf p} \langle \alpha^\dag_{1\mathbf{p}\delta} \beta_{1\mathbf{p} \gamma}  {\sigma}_{\delta \gamma} \rangle$  with momentum  ${\bf Q}_1 $ and $\overrightarrow{ \Delta}_2 \propto \sum_{\bf p} \langle \alpha^\dag_{1\mathbf{p}\delta} \beta_{2\mathbf{p} \gamma}  {\sigma}_{\delta \gamma} \rangle$
with momentum  ${\bf Q}_2 $. Without loss of generality we can
set $\overrightarrow{\Delta}_1$ along 
the $z$-axis and
$\overrightarrow{\Delta}_2$ in the $xz$-plane: 
\begin{eqnarray}
&&\Delta_{1}^{z}= -
U_{SDW}
 \sum_{\bf p} \langle \alpha^\dag_{1\mathbf{p}\uparrow} \beta_{1\mathbf{p} \uparrow}  \rangle \nonumber \\
&&\Delta_{2}^{z(x)}= -
 U_{SDW} \sum_{\bf p} \langle \alpha^\dag_{1\mathbf{p}\uparrow} \beta_{2\mathbf{p}\uparrow(\downarrow)}  \rangle.
\label{f_1_1}
\end{eqnarray}
where $U_{SDW} = U_1 + U_3$.

In the simplest case all masses are equal, {\it i.e.}, $m_x = m_y$ and
$\varepsilon^{\beta_{1}}_{\mathbf{p}} = \varepsilon^{\beta_{2}}_{\mathbf{p}} = \varepsilon^{\beta}_{\mathbf{p}}$, such that 
all circular pockets are perfectly nested. Similarly to the single-band case we can treat the interaction term, Eq.\ref{eq:2}, with the SDW order parameters, Eq.\ref{f_1_1},  
within a mean-field approach. Performing two 
consecutive Bogolyubov transformations\cite{Eremin2010} the quadratic Hamiltonian can be written as
\begin{eqnarray}
\lefteqn{H^{eff}_2 =  \sum_{a,{\bf p}}  \varepsilon^{\beta}_{{\bf p}}  d^\dagger_{a{\bf p}}  d_{a{\bf p}} +} && \nonumber \\
&& \sum_{p} E_{{\bf p}} \left( e^\dagger_{a{\bf p}}  e_{a{\bf p}} + p^{\dagger}_{b{\bf p}}  p_{b{\bf p}} - e^\dagger_{b{\bf p}}  e_{b{\bf p}} - p^{\dagger}_{a{\bf p}}  p_{a{\bf p}}\right),
\label{eq:n7}
\end{eqnarray}
where
$E_{\bf p} =  \pm \sqrt{\left(\varepsilon_{\bf p}\right)^2+|\Delta|^2}$ and $\Delta=\sqrt{(\Delta_1^z)^2+(\Delta_2)^2}$, $\Delta_2=\sqrt{(\Delta_2^z)^2+(\Delta_2^x)^2}$
and there is only one single self-consistent equation for the total gap magnitude
\begin{eqnarray}
 1 = \frac{
U_{SDW}}{2N} \sum_{\bf p} \frac{1}{\sqrt{\left(\varepsilon_{\bf p}\right)^{2}+\Delta^{2}}}.
\label{eq:n2}
\end{eqnarray}

It is clear that the self-consistency equation (\ref{eq:n2}) only fixes the total magnitude of $(\Delta_1^z)^2+(\Delta_2)^2$ but not
the magnitude and the direction of 
 $\Delta_1$ and $\Delta_2$. This implies a huge ground state degeneracy at the mean-field level, where SDW ordering corresponds to 
the spontaneous breaking of an $O(6)$ symmetry. The experimentally realized states with either $\Delta_i=0$ are just two of infinitely many possibilities. Moreover, the degeneracy of the itinerant model is even larger than that in the localized $J_1-J_2$ model where the moments are fixed. In the itinerant picture the magnitude of each sublattice magnetization can 
be different, as long as their sum is kept constant.

In the following we will show that within mean-field the degeneracy can still be lifted by the ellipticity of the electron pockets or 
the interaction within electron or hole pockets. This explains why the particular 
stripe-type order is realized and then due to 
the magneto-elastic coupling the structural order is imposed on the lattice. However, it does not explain why the structural transition occurs sometimes at higher temperature than T$_N$. To cover this aspect we also briefly discuss how the 
Ising-like degeneracy between ${\bf Q}_1$ and ${\bf Q}_2 $ can be lifted
prior to the onset of long-range magnetic ordering, 
giving rise to the so-called Ising-nematic order. We also discuss how this emergent order couples to the
lattice and orbitals.

\subsection{Lifting the magnetic ground state degeneracy at T$_N$}
So far, the analysis assumed the idealized situation of fully nested circular electron and hole pockets with interactions only between holes and electrons. If one takes into account the ellipticity of the electron pockets and additional interactions the degeneracy may be lifted.
W start by first considering the four other possible $\beta-\beta$ interactions:
\begin{eqnarray}
&&H^{ex}_4 =  U_{6} \sum {\beta}^{\dagger}_{1{\bf p}_3 \sigma}
{\beta}^{\dagger}_{2{\bf p}_4 \sigma'}  {\beta}_{2{\bf p}_2 \sigma'} {\beta}_{1{\bf p}_1
\sigma} +  \\ \nonumber
&& U_{7} \sum {\beta}^{\dagger}_{2{\bf p}_3 \sigma}
{\beta}^{\dagger}_{1{\bf p}_4 \sigma'} {\beta}_{2{\bf p}_2 \sigma'} {\beta}_{1{\bf p}_1
\sigma}  \\ \nonumber
&& + \frac{U_{8}}{2} \sum
\left[{\beta}^{\dagger}_{2{\bf p}_3 \sigma} {\beta}^{\dagger}_{2{\bf p}_4 \sigma'}
{\beta}_{1{\bf p}_2 \sigma'} {\beta}_{1{\bf p}_1 \sigma} + h.c \right] + \\ \nonumber
&&\frac{U_{4}}{2} \sum \left[   {\beta}^{\dagger}_{1{\bf p}_3 \sigma}
{\beta}^{\dagger}_{1{\bf p}_4 \sigma'} {\beta}_{1{\bf p}_2 \sigma'} {\beta}_{1{\bf p}_1\sigma} +
  {\beta}^{\dagger}_{2{\bf p}_3 \sigma}
{\beta}^{\dagger}_{2{\bf p}_4 \sigma'} {\beta}_{2{\bf p}_2 \sigma'} {\beta}_{2{\bf p}_1\sigma}  \right]
\label{Hint}
\end{eqnarray}
In the AF state one again applies the sequence of Bogolyubov transformations, 
and takes the appropriate averages $\langle \cdots \rangle $
to obtain the contribution to the ground state energy 
coming from these additional interaction terms. The final correction to the ground state energy
was obtained in \cite{Eremin2010} and has the form:
\begin{eqnarray}
E^{ex}_{gr} & = & 2 A^2 \left[\left(U_{6} + U_8 - U_{7} - U_{4}\right)\right] \frac{|{\Delta}_1|^2 |{\Delta}_2|^2}{\Delta^4} + \nonumber \\
&& 4 A^2 U_{7} \frac{\left({\Delta}_1 \cdot {\Delta}_2 \right)^2} {\Delta^4}
\label{energy}
\end{eqnarray}
Observe that  $E^{ex}_{gr}$ depends on $|{\Delta}_1|^2 |{\Delta}_2|^2$ and
on  $\left({\Delta}_1 \cdot {\Delta}_2 \right)^2$, {\it i.e.},
it is sensitive to both 
the  relative values and relative
directions of ${ \Delta}_1$ and ${\Delta}_2$. When
 all interactions are of equal strength, the first term vanishes, and the
last term favors ${\Delta}_1 \perp { \Delta}_2$.
In this situation,  the $O(6)$ degeneracy of the 
perfect-nesting model model is broken, but
only down to $O(3) \times 
O(3)$, {\it i.e.},
the magnitude of the order parameter at each site  is now the same because
 $\left(\overrightarrow{ \Delta}_1+\overrightarrow{\Delta}_2\right)^2=\left(\overrightarrow{ \Delta}_1-\overrightarrow{ \Delta}_2\right)^2$, but
the angle between the directions of the SDW order in the two sublattices
(i.e., between $\overrightarrow{\Delta}_1 +\overrightarrow{\Delta}_2$ and $\overrightarrow{\Delta}_1
-\overrightarrow{\Delta}_2$) is still arbitrary. This is exactly the same situation as
in the classical $J_1-J_2$ model.  However, once $U_6+ U_8-U_7-U_4$ is nonzero, the degeneracy is broken down to a conventional $O(3)$ already at the mean-field level.
Because $U_4$ is reduced and even changes sign under 
the RG flow
~\cite{Chubukov2008}, while other $U_i$ do not flow, the most likely situation is that  $U_6+ U_8-U_7-U_4 >0$, in which case $E^{ex}_{gr}$ is minimized when
either $\overrightarrow{\Delta}_1 =0$, or  $\overrightarrow{\Delta}_2 =0$, i.e., SDW order is either $(0,\pi)$ or $(\pi,0)$. This is exactly the same SDW order as observed in experiments.  If  $U_6+ U_8-U_7-U_4$ was negative,
$E^{ex}_{gr}$  would be minimized when  $|\overrightarrow{\Delta}_1| =|\overrightarrow{\Delta}_2|$, in which case the SDW OPs of the two  sublattices would align orthogonal to each other. The spin configuration  for such state is shown 
in the left panel of Fig.\ref{MagOrdRealSpace}.

We also consider the impact of the elliptical distortion of the electron Fermi pockets to the enlarged $O(6)$ symmetry of the perfect-nesting model. 
The effective masses $m_x$ and $m_y$ are not equal, and  $\varepsilon^{\beta_1}_{{\bf k}} \neq \varepsilon^{\beta_2}_{{\bf k}}$. To continue with the analytical analysis, one assumes that the ellipticity is small, 
introduces $m_x = (1+\delta)m$ and $m_y = (1-\delta)m$, where  $\delta <<1$,
and computes the correction to the ground state energy to second order in $\delta$.
The contribution to the ground state energy coming from the ellipticity is
\begin{eqnarray}
E_{gr}^{ellipt} = C |\overrightarrow{\Delta}_1|^2 |\overrightarrow{\Delta}_2|^2,  ~~C =\delta^2 \frac{m \mu^2}{4 \pi \Delta^4}
\label{su_6}
\end{eqnarray}
where the coefficient $C$ is positive, i.e., the correction due to 
the ellipticity of the electron pockets again breaks the degeneracy and selects either 
the $(0,\pi)$ or $(\pi,0)$ state.
This is again the same 
magnetic order that is observed experimentally.
It is remarkable that ellipticity leads to a term in the ground state energy that is similar to an effective interaction between two SDW OPs which, for $\Delta << \mu$,
leads to the same selection of the ground state SDW order as the direct interaction between the two electron pockets.

\subsection{Ising nematic order above $T_N$}

The onset of magnetic order with ordering vector $(0,\pi)$ or $(\pi,0)$ breaks not only the $O(3)$ spin-rotational symmetry, but also the $C_4$ tetragonal symmetry of the lattice. Experimentally, this tetragonal symmetry breaking has however been observed at temperatures $T_s$ larger than $T_N$ for many compounds\cite{Fisher2011}. In the temperature regime between $T_s$ and $T_N$, the system displays not only an orthorhombic distortion, but also in-plane anisotropies in several observables, such as dc \cite{Chu2010,Tanatar2010} and ac conductivities  \cite{Dusza2011,Nakajima2011}, as well as the magnetic susceptibility \cite{Matsuda2012}. Because this phase breaks the rotational symmetry of the lattice without affecting its translational symmetry, it has been called the Ising-nematic phase.

\begin{figure}[h]
\begin{center}
\includegraphics[width=7cm]{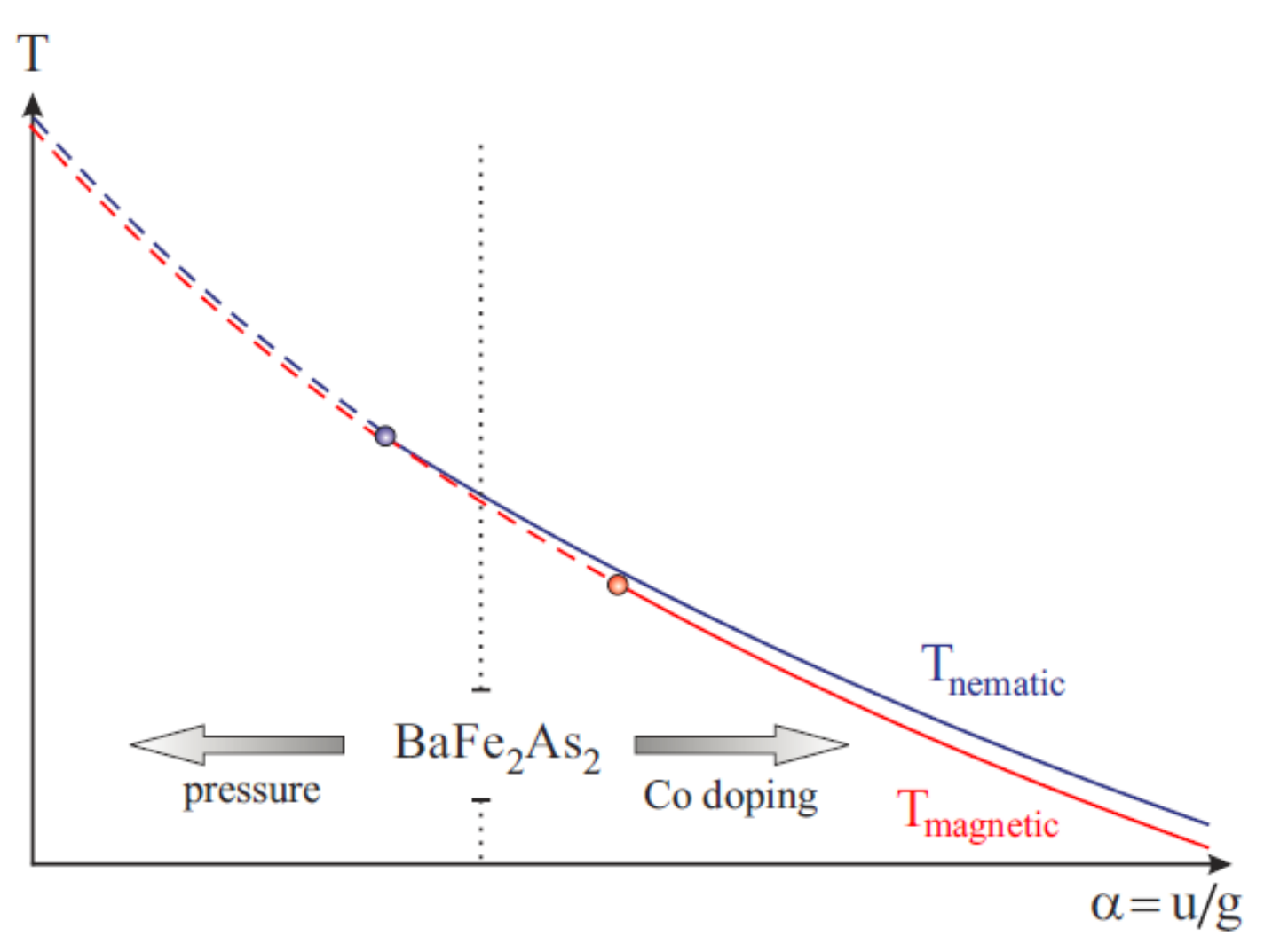}
\end{center}
\caption{(Color online) Characteristic phase diagram as function of temperature and inverse nematic coupling
for 
moderately anisotropic, quasi-two dimensional
systems\cite{Fernandes2011}. Red and blue curves represent
magnetic and Ising-nematic transitions, respectively.
We show schematically how application of pressure or electron doping (via Co substitution) changes the system behavior of BaFe$_2$As$_2$.
Here a solid (dashed) line denotes a second-order
(first-order) transition, and a double-dashed line indicates a
simultaneous first-order transition. The two solid points mark
the positions of the nematic and magnetic tri-critical points.}
\label{fig:nematic}
\end{figure}

To discuss how an Ising-nematic phase appears in our model, we first note that the order parameter manifold of the ground state is $O(3)\times Z_2$, where the $Z_2$ symmetry refers to selecting one of the two ordering vectors $(0,\pi)$ or $(\pi,0)$. Thus, if the $Z_2$ symmetry is broken the system is no longer tetragonal, since the $x$ and $y$ directions become different inside the Fe-square unit cell. Although as discussed above both $O(3)$ and $Z_2$ symmetries are broken simultaneously in the mean-field level, fluctuations can split their transition temperatures. Then, the intermediate phase where the $Z_2$ symmetry is broken but the spin-rotational $O(3)$ symmetry is kept intact corresponds to the Ising-nematic phase \cite{Fang2008,Xu2008,Qi2009,Fernandes2010}. This phase is known to appear in the localized  $J_1-J_2$ model \cite{Chandra1990,Si2008}, and we now review its origin within our itinerant fermonic model \cite{Fernandes2011}. We start from the microscopic Hamiltonian  Eq.\ref{eqH} and Eq.\ref{eq:2} and write down the partition function
\begin{equation}
Z\propto\int d\overrightarrow{\Delta}_{1}d\overrightarrow{\Delta}_{2}\mathrm{e}^{-S_{\mathrm{eff}}\left[\overrightarrow{\Delta}_{1},\overrightarrow{\Delta}_{2}\right]}
\label{aux_action_1}
\end{equation}
in terms of the  effective action for the two magnetic order parameters
\begin{eqnarray}
S_{\mathrm{eff}}\left[\overrightarrow{\Delta}_{1},\overrightarrow{\Delta}_{2}\right] & = & r_{0}\left(\overrightarrow{\Delta}_{1}^{2}+\overrightarrow{\Delta}_{2}^{2}\right)+\frac{u}{2}\left(\overrightarrow{\Delta}_{1}^{2}+
\overrightarrow{\Delta}_{2}^{2}\right)^{2}\nonumber \\
 &  & -\frac{g}{2}\left(\overrightarrow{\Delta}_{1}^{2}-\overrightarrow{\Delta}_{2}^{2}\right)^{2}+v\left(\overrightarrow{\Delta}_{1}\cdot\overrightarrow{\Delta}_{2}\right)^{2}.
 \label{action}
 \end{eqnarray}
The coefficients can be computed in terms of the bare fermionic propagators 
and give $u>0$, $g>0$, and $v=0$ for an expansion near perfect nesting, with $g \propto \delta^2$, in agreement with the results of the previous section. Indeed, mean-field minimization of this free energy yields the state $\langle\overrightarrow{\Delta}_i\rangle \neq 0$ with either $i=1$ or $i=2$.
The  $Z_2$ symmetry breaking
takes place when fluctuations associated with one
of the bosonic fields are larger than the fluctuations associated with the other one, e.g.,
$\langle\overrightarrow{\Delta}_1^2\rangle > \langle\overrightarrow{\Delta}_2^2\rangle$ while $\langle\overrightarrow{\Delta}_1\rangle=\langle\overrightarrow{\Delta}_1\rangle=0$.
To capture this physics, we introduce the scalar fields $\phi=\langle\overrightarrow{\Delta}_1^2\rangle - \langle\overrightarrow{\Delta}_2^2\rangle$ and $\psi=\langle\overrightarrow{\Delta}_1^2\rangle + \langle\overrightarrow{\Delta}_2^2\rangle$ via two Hubbard-Stratonovich transformations. Note that in contrast to the previous subsection we do not treat $\overrightarrow{\Delta}$ as a mean-field order but rather as a fluctuating quantity. Here the field $\phi$ is associated with the appearance of Ising-nematic order. 
Focusing on the paramagnetic phase, we integrate out the  $\overrightarrow{\Delta}$ fields and obtain an effective action in terms of the new scalar fields only:
\begin{equation}
S_{\mathrm{eff}}\left[\phi,\psi\right]=\int_{q}\left\{ \frac{\phi^{2}}{2g}-\frac{\psi^{2}}{2u}+\frac{3}{2}\log\left[\left(\chi_{q}^{-1}+\psi\right)^{2}-\phi^{2}\right]\right\} \label{action_phi}
\end{equation}
The action can now be analyzed 
within the saddle-point approximation, which exact in the limit where the number of components $N$ of the $\overrightarrow{\Delta}$ field is large \cite{Fernandes2011}.
The result is that, for quasi-2D systems, the splitting and the characters of the magnetic and nematic transitions depend on the dimensionless inverse nematic coupling $\alpha=\frac{u}{g}$, which is found to
decrease with pressure but increase with electron doping (see Fig.\ref{fig:nematic}).

As shown in Fig. \ref{fig:nematic}, from \cite{Fernandes2011}, depending on the strength of the coupling $\alpha$ different types of system behavior can be found.
In the case of quasi-2D systems,
for large values of the nematic coupling (small $\alpha$), magnetic order emerges simultaneously to the Ising-nematic order, via a first-
order transition ($T_N = T_s$).  For intermediate couplings  $\alpha$, 
a second-order Ising-nematic transition 
is followed by a meta-nematic transition, in which the nematic order parameter jumps between two non-zero values. The feedback of nematic order on the magnetic spectrum causes a first-order magnetic transition simultaneously to the meta-nematic transition, but split from the second-order nematic transition. This sequence of transitions agrees with the ones observed experimentally for BaFe$_2$As$_2$ \cite{Matsuda2012}.
For small nematic couplings (large $\alpha$) a second-order Ising-nematic transition is followed by a second-order magnetic stripe transition at a smaller temperature. Within our model, doping electrons into the system decreases the nematic coupling, driving the system to this regime of split second-order transitions. Experimentally, this sequence of transitions is observed in Co-doped BaFe$_2$As$_2$ compounds, which display more negative carriers than the undoped parent compound.

Due to the magneto-elastic coupling and the coupling to the $d_{xz}$ and $d_{yz}$ Fe-orbital degrees of freedom, the onset of Ising-nematic order simultaneously triggers a non-zero orthorhombic distortion as well as ferro-orbital order \cite{Fernandes2011}. Indeed, a splitting in the onsite energies of the $d_{xz}$ and $d_{yz}$ orbitals has been observed experimentally \cite{Yi2011}, and the sign of the splitting agrees with the one predicted by the itinerant model. We note that ferro-orbital order has also been discussed within approaches that focus on a spontaneous breaking of the $d_{xz}$ and $d_{yz}$ orbital symmetry (see for instance \cite{Kruger2009,Lv2010,Chen2010}).


\section{Quasiparticle Interference in the SDW state: possible multiorbital effects}\label{sec1.3}

So far we have not discussed the role of orbitals in the formation of the AF order. At the same time,
there is no doubt that the orbitals do play some role due to the fact that all 5 
Fe-$3d$ orbitals contribute to the bands, crossing the Fermi level. For 
instance, they introduce a significant
angular dependence for the intraband and interband interactions
resulting in a change of important details in the
superconducting and antiferromagnetic states. It was argued that the originally nodeless s$^{+-}$-wave superconducting
state may acquire nodes on the electron pockets 
due to this angular dependence.\cite{Kuroki2009,Wang2010,Maiti2011,Thomale2011,Kemper2010}  For the AF state it was predicted that iron-based superconductors show nodal elementary excitations with a linear Dirac spectrum at low energies as a result of orbital effects\cite{Vishwanath2009,Yang2010,Morinari2010}.  In addition, due to the fact that the antiferromagnetic order is either $(\pi,0)$ or $(0,\pi)$, the magnetic anisotropy induces orbital anisotropies which in turn affect the electronic properties along $a$ and $b$ directions\cite{Sugimoto2011,Daghofer2010,Valenzuela2010}.
They may also introduce other ordered states, such as ferro-orbital order\cite{Thalmeier2008,Lv2010} or antiferro-orbital order\cite{Kontani2012}. As discussed above, ferro-orbital order was observed experimentally and naturally appears within our model for the magnetic instability.
On the other hand, antiferro-orbital order, which has not been observed experimentally to the best of our knowledge, certainly contradicts the concept of purely magnetically driven instabilities.

Within this multi-orbital scenario, the AF order may connect similar or different orbitals. To distinguish these two cases experimentally, and shed light on the nature of the AF state, one may investigate how the electronic spectrum is changed in each case. One way to probe the electronic spectrum is via quasiparticle interference (QPI), as measured by scanning tunneling microscopy. Previously, QPI in the AF state of the ferropnictides was discussed by several groups experimentally and theoretically.\cite{Chuang2010,Knolle2010b,Akbari2010,Plonka2013,Rosenthal13}  In the following we discuss the 
signatures in the QPI of different forms of the AF order parameter in orbital space. 

\subsection{Two orbital model}

In order to 
gain insight on the formalism, and 
in particular to see the effect of the orbital degrees of freedom, we concentrate on the simplest two-orbital model that was introduced in the context of FPs \cite{Raghu2008}. It takes into account only the $d_{xz}$ and $d_{yz}$ orbitals which, also experimentally and from LDA calculations, have the largest weight around the Fermi energy.

The kinetic part of the Hamiltonian is written in the two component spinor basis in orbital space with the help of the Pauli matrices $\tau_i$
\begin{eqnarray}
\label{Hraghu}
H_0 & = &  \sum_{{\bf k},\sigma }  \psi_{\sigma}^{\dagger} \left[ \left( \epsilon_{+} -\mu \right) \tau_0 + \epsilon_{-}  \tau_3 +\epsilon_{xy}  \tau_1\right]  \psi_{\sigma}
\end{eqnarray}
where
\begin{eqnarray}
\label{Spinor}
 \psi_{\sigma}({\bf k}) & = &
\begin{pmatrix}
d_{xz \sigma} ({\bf k}) \\
d_{yz \sigma} ({\bf k})
\end{pmatrix}.
\end{eqnarray}
Here, $\varepsilon_\pm(k)  =  \frac{\varepsilon_x(k)\pm\varepsilon_y(k)}{2}$, $\varepsilon_x(k)  =  -2t_1\cos k_x-2t_2\cos k_y-4t_3\cos k_x\cos k_y$, $\varepsilon_y(k) =  -2t_2\cos k_x-2t_1\cos k_y-4t_3\cos k_x\cos k_y$, and
$\varepsilon_{xy}(k)  =  -4t_4\sin k_x\sin k_y$.

In general, the interacting part of the Hamiltonian consists of different on-site electron-electron interactions:
\begin{eqnarray}
\label{HraghuInt}
H^{orb}_{int} & = &  U \sum_{i } \sum_{\nu} n_{i\nu\uparrow}n_{i\nu\downarrow} +V \sum_{\nu \neq \mu,\sigma,\sigma'}
n_{i\nu\sigma}n_{i\mu\sigma'} - \\
 & & J \sum_{\nu \neq \mu} {\bf S}_{i\nu}\cdot {\bf S}_{i\mu} +J^{\prime}\sum_{\nu\neq\mu} d_{i\nu\uparrow}^{\dagger} d_{i\nu\downarrow}^{\dagger}d_{i\mu\downarrow}d_{i\mu\uparrow},
\end{eqnarray}
where $U$ and $V$ refer to the intra- and inter-orbital Coulomb repulsion, $J$ and $J^{\prime}=J/2$ denote the Hund and the pair hopping terms.

\subsection{Quasiparticle interference in the SDW state.}

The particular way in which the system reacts to 
an impurity can be used as a probe for the underlying nature of the many body state.
Our main goal is to use quasiparticle interference (QPI) to detect subtle multi-orbital effects in the SDW state. Therefore, we 
consider not only intra-orbital magnetism, but also the possibility of inter-orbital magnetism and orbital order. We 
set the moment to be along the $z$ direction and study the following mean-field Hamiltonian matrix:

\begin{eqnarray}
& \hat h^{Orbital}(\mathbf{k},\sigma)  = \\ \nonumber
& \begin{pmatrix}
e_x(\mathbf{k})+\eta & e_{xy}(\mathbf{k}) & \sigma M_x^{||} & \sigma M^{\perp} \\
e_{xy}(\mathbf{k}) & e_y(\mathbf{k})-\eta & \sigma M^{\perp} & \sigma M_y^{||} \\
\sigma M_x^{||} & \sigma M^{\perp} &  e_x(\mathbf{k+Q}) +\eta&  e_{xy}(\mathbf{k+Q}) \\
\sigma M^{\perp} & \sigma M_y^{||} &  e_{xy}(\mathbf{k+Q}) &  e_y (\mathbf{k+Q}) -\eta
\end{pmatrix}
\end{eqnarray}

Here, $M_i^{||}$ is the intra orbital magnetic order parameter of the $i^{\text{th}}$ orbital and $M^{\perp}$ is the inter-orbital magnetic order parameter between the $xz$ and $yz$ orbitals. 
We consider the ordering wave vector to be $(\pi,0)$. The parameter $\eta$ describes the orbital splitting of the otherwise degenerate $xz$, $yz$ orbitals due to ferro-orbital ordering. 
The two magnetic order parameters and the ferro-orbital order could be computed within 
a mean-field approximation using an appropriate decoupling of Eqs.(15)-(16) but as we are only interested in qualitative features of the QPI maps we use them as parameters.

To proceed, we add the impurity term to the Hamiltonian
\begin{eqnarray}
\label{Himp}
H_{\text{Imp}} & = & \sum_{\mathbf{k, k'},\sigma, \sigma', i, j}\underbrace{\left(  U^{i,j} \delta_{\sigma,\sigma'} + J^{i,j} \tau^{z}_{\sigma,\sigma'}\right)}_{\equiv \hat V_{\sigma}} d^{\dagger}_{i \sigma} ({\bf k}) d_{j \sigma'} ({\bf k'})
\end{eqnarray}
in which $U^{i,i}$ is a local potential scatterer within the $i^{\text{th}}$ orbital, 
and $U^{i,j}$ with $i \neq j$ corresponds to inter-orbital scattering that is considered to be weak.
Furthermore, we take into account a classical magnetic impurity oriented in the $z$-direction and parametrized by $J^{i,j}$ (indices have a similar meaning as before).

The full Green's function is calculated via
\begin{eqnarray}
\label{Gfull}
\hat G_{\sigma} (\mathbf{k,k'},\omega) & = & \hat G^0_{\sigma} (\mathbf{k},\omega) \delta_{\mathbf{k,k'}} + \hat G^0_{\sigma} (\mathbf{k},\omega) \hat T_{\sigma}(\omega) \hat G^0_{\sigma} (\mathbf{k'},\omega)
\end{eqnarray}
with the bare Green's function (GF)
\begin{eqnarray}
\label{GFbare}
\hat G^0_{\sigma} (\mathbf{k},\omega) & = & \left( \omega  - \hat h^{Orbital}(\mathbf{k},\sigma)\right)^{-1}
\end{eqnarray}
and the energy dependent T-matrix
\begin{eqnarray}
\label{Tmatrix}
\hat T_{\sigma} (\omega) = \left( 1- \hat V_{\sigma} \frac{1}{N} \sum_{\mathbf{p}} \hat G^0_{\sigma} (\mathbf{p},\omega) \right)^{-1} \hat V_{\sigma}.
\end{eqnarray}

From 
these expressions, the local density of states is calculated as
\begin{eqnarray}
\label{LDOS}
\rho_{\sigma}(\mathbf{r},\omega) & = & -\frac{1}{2 \pi \sqrt{N} } \text{Im} \sum_{\mathbf{q}}  e^{i \mathbf{r q}} N_{\sigma}(\mathbf{q},\omega) \\
\text{with} \ \  N_{\sigma}(\mathbf{q}, \omega) & = & \frac{1}{\sqrt{N}} \sum_{\mathbf{k}} \text{tr} \hat G_{\sigma} (\mathbf{k,k+q}, \omega)
\end{eqnarray}
and the trace is taken over the orbital and sublattice indices.
In quasiparticle interference we are actually only interested in the impurity induced ${\mathbf q}$-dependent interference contribution to the local density of states which is given by
\begin{eqnarray}
\label{QPIdos}
\delta N_{\sigma}(\mathbf{q},\omega) & = & \frac{1}{\sqrt{N}} \sum_{\mathbf{k}} \text{tr} \hat G^0_{\sigma} (\mathbf{k},\omega) \hat T_{\sigma}(\omega) \hat G^0_{\sigma} (\mathbf{k+q},\omega) \nonumber \\
\end{eqnarray}

\begin{figure}
\begin{centering}
\includegraphics[width=0.8\columnwidth]{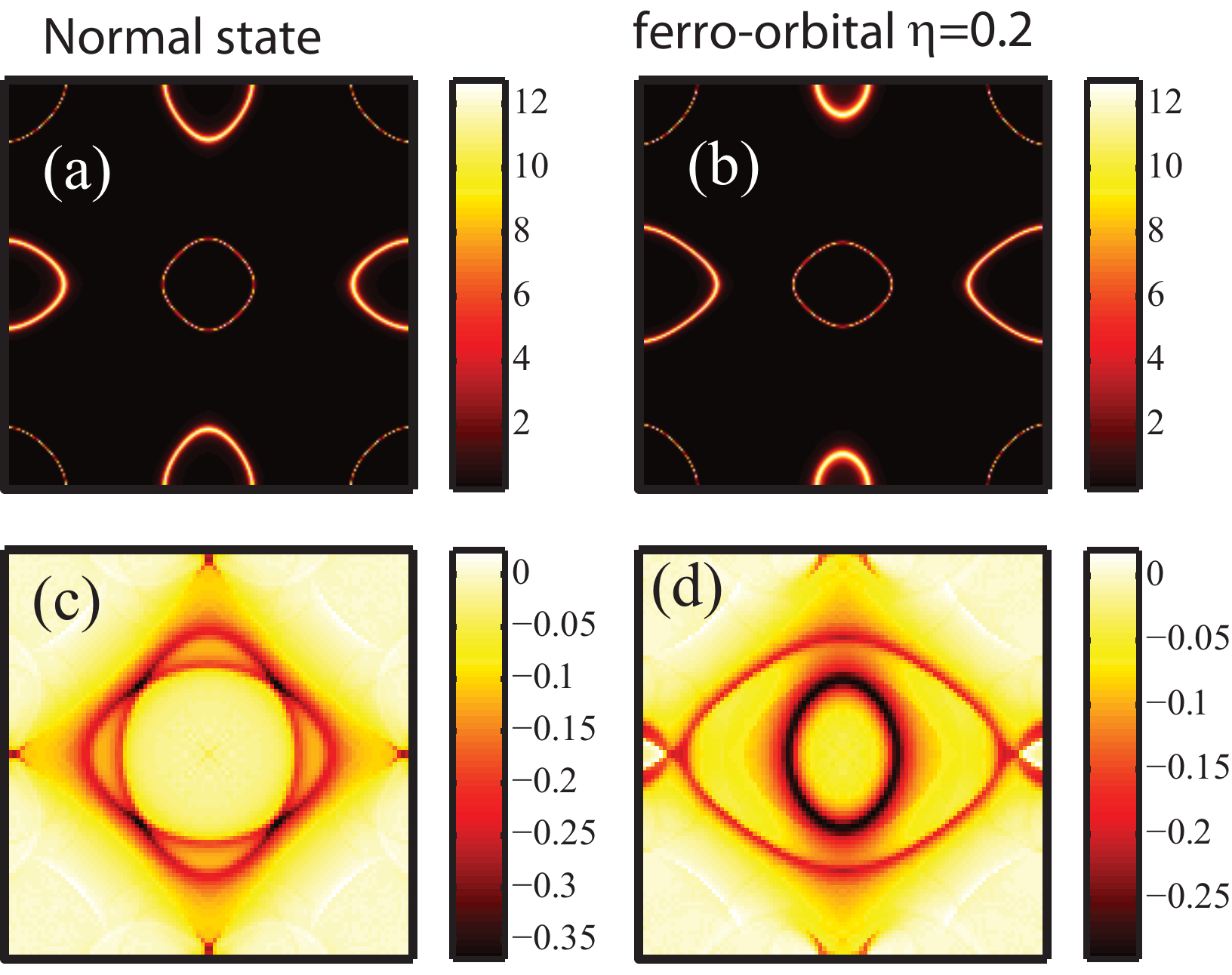}
\par\end{centering}
\caption{(Color online) Comparison 
between the normal state (left) and the nematic state with an orbital splitting $\eta=200 meV$ (right) for an energy of $\omega=10 meV$. 
Upper (lower) panels refer to the spectral function (QPI pattern). \label{Comparison1}}
\label{NormalVsAn}
\end{figure}

In Fig.\ref{NormalVsAn} we compare the paramagnetic state without (left column) and with 
ferro-orbital order ($\eta=0.2$eV, right column). Already in the bare spectral function at the Fermi energy (panels (a) and (b)) the reduced $C_2$ symmetry of the state with orbital ordering is apparent. It is even more pronounced in the QPI signal $\delta N_{\sigma}(\mathbf{q},\omega)$ (panels (c) and (d)), which can be used as an experimental probe to detect this ferro-orbital order.

\begin{figure}
\begin{centering}
\includegraphics[width=0.8\columnwidth]{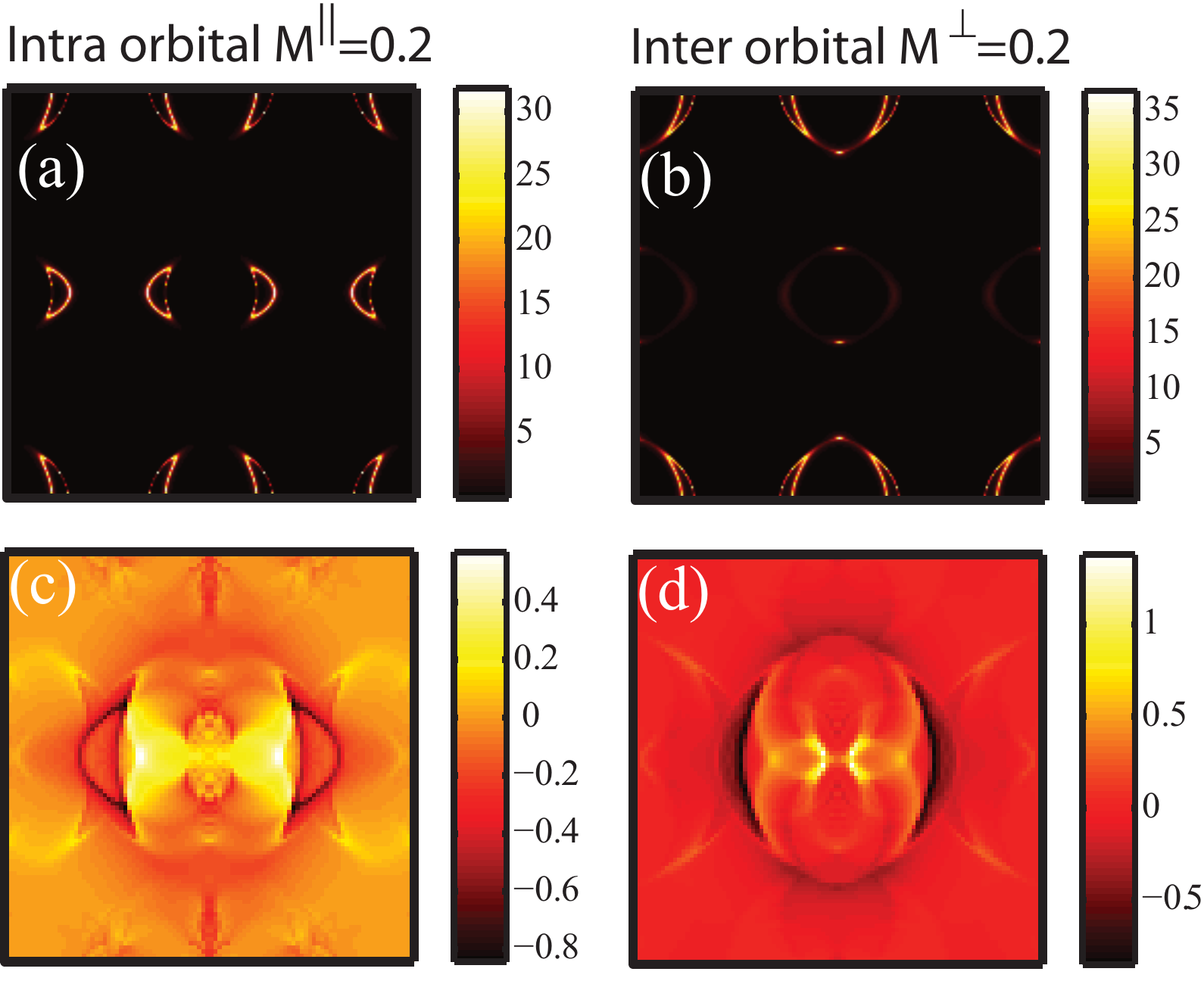}
\par\end{centering}
\caption{(Color online) Comparison 
between intra (left) and inter (right) orbital magnetism, for a magnetic gap of $200meV$ and for an energy of $10 meV$. 
Upper (lower) panels refer to the spectral function (QPI pattern). \label{Comparison2}}
\label{InterVsIntra}
\end{figure}

Next, we move to the magnetic state to discuss whether 
it is possible to distinguish magnetism arising mostly from inter-orbital ($M^{\perp}$) or intra-orbital ($M^{||}$) contributions.
In Fig.\ref{InterVsIntra} we show again the spectral function (first row) and the QPI signal (second row) for the two different magnetic
scenarios with a magnetic gap of $~200$meV in each case. Since 
the parts of the Fermi surface that are connected by the antiferro-magnetic wave vector $\mathbf{Q}$ are mostly of the same orbital character, the Fermi surface has much more pronounced gap openings in the intra-orbital magnetic scenario. Note that this is the simple reason why in a self-consistent mean-field treatment this state dominates. In the QPI signal the two scenarios of intra- and inter-orbital magnetism can be clearly distinguished.
In particular, the main difference between the intraorbital and the interorbital antiferromagnetism is related to the character of the band reconstruction in the corresponding AF states.  In the case of intraorbital AF state the resulting constant energy cuts show characteristic small pockets near the $\Gamma-$point of the BZ that are quite stable against the increasing magnitude of the AF gap. The stability of these pockets is related to the nodal structure of the intraorbital AF order parameter in the momentum space  as was discussed previously\cite{Vishwanath2009}. Therefore, the QPI pattern in this case shows characteristic structure at low energies which directly reflects the  structure of the pockets in the spectral function (see Fig.\ref{InterVsIntra}(a),(c)). In addition, observe that the resulting pockets in the spectral function are almost identical around the $\Gamma-$point and
$(0,\pi)$ points of the BZ. These points are not necessarily equivalent in the case of $(\pi,0)$ magnetic order but they are indeed equivalent in the case of intraorbital AF order and if one neglects weak ferrorbital order. This additional symmetry is related to the specific situation of the intraorbital order as the folding of the hole pocket at $(0,0)$ and electron pocket at $(\pi,0)$ is driven by the gap in the $yz$-orbital, while equal size gap in the $xz$-orbital drives the reconstruction of the
hole pocket at $(\pi,\pi)$ and the electron pocket at $(0,\pi)$. As both gaps come out equal in the calculation  and because the orbital content of each pocket interchanges by adding $(\pi,\pi)$ momentum the resulting band structure topology shows this extra symmetry. Note that due to additional non-zero small ferrorbital order the pockets are not completely equivalent around the $\Gamma$-point and $(0,\pi)$ points but still very similar as seen from Fig.5(a) .

In the case of interorbital AF order this symmetry is lost and overall the band structure reconstruction is quite different. For example, as shown in Fig.\ref{InterVsIntra}(b),(d) the spectral function is very different between $(0,0)$ and $(0,\pi)$ point of the BZ. In addition, the reconstruction in the interorbital AF state does not show small size pockets which are characteristic of the intraorbital AF state. These main differences should be also preserved in a more sophisticated 5 and 10-orbital models and may serve for the experimental identification of the origin of the AF state in ferropnictides.

\section{Discussions and conclusions}
In this 
paper we reviewed the itinerant description of antiferromagnetism in parent materials of the iron-based superconductors. In contrast to a purely localized scenario this theory allows for a coherent understanding of the full phase diagram of these materials as a function of doping, disorder, and pressure. All the  magnetic properties such as magnitude of the magnetic moment, selection of the order from the degenerate manifold of the possible ground states, 
and appearance of the Ising-nematic order above T$_N$ can be well understood and connected to the basic electronic structure of these materials.

In principle one can draw a phenomenological connection between the
itinerant model and the $J_1- J_2$ model. In the latter, one has to replace the soft constraint
on the ``momentum-space'' magnetic order parameters $\boldsymbol{\Delta}_{i}$ by a hard constrain on
the ``real-space'' magnetic order parameters $\mathbf{M}_{i}$. These order parameters correspond to the
magnetizations of the two weakly-coupled interpenetrating Neel sublattices that constitute the stripe
magnetic configuration. They are related to $\boldsymbol{\Delta}_{i}$ by $\boldsymbol{\Delta}_{1}=\mathbf{M}_{1}+\mathbf{M}_{2}$
and $\boldsymbol{\Delta}_{2}=\mathbf{M}_{1}-\mathbf{M}_{2}$. Furthermore, in the $J_1- J_2$ model one finds
$g \propto J_{1}^{2}/J_{2}$, which in generally is small $g\ll J_2$ \cite{Chandra1990}, and gives rise to a pre-emptive Ising-nematic order
that breaks the $Z_{2}$ degeneracy associated with the stripe magnetic configuration.
Therefore, in the localized scenario,  $g$ is generally small and insensitive to doping, in contrast to the itinerant scenario (see Fig. \ref{fig:nematic}), whose phase diagram accounts for the experimentally observed evolution of the transitions in BaFe$_2$As$_2$.
Furthermore, one has to be careful with a description of the spin waves in iron-based superconductors  via $J_1 -J_2$ like models.
The original derivation of the $J_1-J_2$ model goes back to the single-band Hubbard model with hopping between nearest and next-nearest neighbors. Iron-based superconductors are 
multi-orbital materials and their localized strong-coupling limit is not exactly known at present. Therefore, spin wave calculations based on the $J_1-J_2$ model even taking into account biquadratic couplings have limitations especially at higher energies.

Nevertheless, there are also limitations of the itinerant description at present. 
For instance, in the iron chalcogenides $\mathrm{FeTe_{1-x}Se_{x}}$, the itinerant scenario is applicable in the regime of intermediary Se doping, near
the superconducting dome of the $\left(x,T\right)$ phase diagram.
In this region, the electronic structure is similar to the one considered in 
our itinerant model and neutron scattering
shows that magnetic fluctuations are peaked at the ordering vectors
$\mathbf{Q}_{1}=\left(\pi,0\right)$ and $\mathbf{Q}_{2}=\left(0,\pi\right)$.
On the other hand, the itinerant model fails for 
the undoped sample where nesting might be absent and the magnetic order has
$(\pi/2,\pi/2)$ ordering wave vector. The same problem concerns $A_x$Fe$_{2-x/2}$Se$_2$ [$A=$Cs, K, (Tl,Rb), (Tl,K)] iron-selenide compounds
where the electronic structure is still discussed.

Further open issues on the theoretical side which have to be studied are the possible role of  spin-orbit coupling, interaction with magnetic and non-magnetic impurities, and the doping dependence 
of magnetism. This promises interesting perspectives for future research in the field of iron-based superconductors.

We acknowledge helpful discussions and collaborations with A. Akbari, Ph. Brydon, P.
Hirschfeld, S. Maiti, Y. Matsuda, R. Moessner,  P.Thalmeier, R. Thomale, and M. Vavilov.
IE acknowledges financial support of the SPP 1458 'Eisen-Pniktide' of the Deutsche Forschungsgemeinschaft, the German Academic Exchange Service (DAAD PPP USA No. 57051534), and the Mercator Research Center Ruhr. J.K.~acknowledges support from the Studienstiftung des deutschen Volkes, the IMPRS Dynamical Processes in Atoms, Molecules and Solids, DFG within GRK1621.


\begin{thebibliography}{99}

\bibitem{kamihara2008} Y. Kamihara, T. Watanabe, M. Hirano, and H. Hosono, J. Am. Chem. Soc. {\bf 130} 3296 (2008).

\bibitem{mazin} P. J. Hirschfeld, M. M. Korshunov, I. I. Mazin, Rep. Prog. Phys. {\bf 74}, 124508 (2011).

\bibitem{Kaminski2008} C. Liu, G. D. Samolyuk, Y. Lee, N. Ni, T. Kondo, A. F. Santander-Syro, S. L. Bud'ko, J. L. McChesney, E. Rotenberg, T. Valla, A. V. Fedorov, P. C. Canfield, B. N. Harmon, A. Kaminski, Phys. Rev. Lett. {\bf 101}, 177005 (2008).

\bibitem{Terashima2009} K. Terashima, Y. Sekiba, J. H. Bowen, K. Nakayama, T. Kawahara, T. Sato, P. Richard, Y.-M. Xu, L. J. Li, G. H. Cao, Z.-A. Xu, H. Ding, T. Takahashi, Proc. Natl. Acad. Sci. {\bf 106}, 7330 (2009).

\bibitem{Borisenko2009} V.B. Zabolotnyy, D.S. Inosov, D.V. Evtushinsky, A. Koitzsch, A. A. Kordyuk, G. L. Sun, J. T. Park, D. Haug, V. Hinkov, A. V. Boris, C. T. Lin, M. Knupfer, A. N. Yaresko, B. Buechner, A. Varykhalov, R. Follath, S. V. Borisenko, Nature {\bf 457}, 569 (2009).

\bibitem{Feng2009} L. X. Yang, Y. Zhang, H. W. Ou, J. F. Zhao, D. W. Shen, B. Zhou, J. Wei, F. Chen, M. Xu, C. He, Y. Chen, Z. D. Wang, X. F. Wang, T. Wu, G. Wu, X. H. Chen, M. Arita, K. Shimada, M. Taniguchi, Z. Y. Lu, T. Xiang, D. L. Feng, Phys. Rev. Lett. {\bf 102}, 107002 (2009).

\bibitem{Shen2008} D. H. Lu, M. Yi, S.-K. Mo, A. S. Erickson, J. Analytis, J.-H. Chu, D. J. Singh, Z. Hussain, T. H. Geballe, I. R. Fisher, Z.-X. Shen, Nature {\bf 455}, 81 (2008).

\bibitem{Ding2011} H. Ding, K. Nakayama, P. Richard, S. Souma, T. Sato, T. Takahashi, M. Neupane, Y.-M. Xu, Z.-H. Pan, A.V. Federov, Z. Wang, X. Dai, Z. Fang, G.F. Chen, J.L. Luo, N.L. Wang, J. Phys.: Condens. Matter {\bf 23}, 135701 (2011); P. Richard, T. Sato, K. Nakayama, T. Takahashi, H. Ding, Rep. Prog. Phys. {\bf 74}, 124512 (2011).

\bibitem{Coldea2008} A. I. Coldea, J. D. Fletcher, A. Carrington, J. G. Analytis, A. F.
Bangura, J.-H. Chu, A. S. Erickson, I. R. Fisher, N. E. Hussey,
and R. D. McDonald, Phys. Rev. Lett. {\bf 101}, 216402 (2008); H. Shishido, A.F. Bangura, A.I. Coldea, S. Tonegawa, K. Hashimoto, S. Kasahara, P.M.C. Rourke, H. Ikeda, T. Terashima, R. Settai, Y. Onuki, D. Vignolles, C. Proust, B. Vignolle, A. McCollam, Y.Matsuda, T. Shibauchi, A. Carrington,
Phys. Rev. Lett. {\bf 104}, 057008 (2010).

\bibitem{Carrington} A Carrington, Rep. Prog. Phys. {\bf 74} 124507 (2011).

\bibitem{Singh2008} D.J. Singh and M.-H. Du, Phys. Rev. Lett. {\bf 100}, 237003 (2008).

\bibitem{Boeri2008} L. Boeri, O.V. Dolgov, and A. A. Golubov, Phys. Rev. Lett. {\bf 101}, 026403
(2008).


\bibitem{Fernandes2011} R.M. Fernandes, A.V. Chubukov, J. Knolle, I. Eremin, J. Schmalian, Phys. Rev. B {\bf 85}, 024534 (2011).

\bibitem{Eremin2010} I. Eremin and A.V. Chubukov, Phys. Rev. B {\bf 81}, 024511 (2010).


\bibitem{Chandra1990} P. Chandra, P. Coleman and A.I. Larkin, Phys. Rev. Lett. {\bf 64}, 88 (1990).

\bibitem{Si2008} Q. Si and E. Abrahams, Phys. Rev. Lett. {\bf 101}, 076401 (2008).

\bibitem{Xu2008} Cenke Xu, Markus Muller, and Subir Sachdev, Phys. Rev. B 78, 020501(R) (2008).

\bibitem{Yildirim2008} T. Yildirim, Phys. Rev. Lett. {\bf 101}, 057010 (2008).

\bibitem{Uhrig2009} Goetz S. Uhrig, Michael Holt, Jaan Oitmaa, Oleg P. Sushkov, and Rajiv R. P. Singh, Phys. Rev. B {\bf 79}, 092416 (2009).

\bibitem{Uchida2009} M. Nakajima, S. Ishida, K. Kihou, Y. Tomioka, T. Ito, Y. Yoshida, C. H. Lee, H. Kito, A. Iyo, H. Eisaki, K. M. Kojima, and S. Uchida, Phys. Rev. B {\bf 81}, 104528 (2010).

\bibitem{Rice1970} M.T. Rice, Phys. Rev. B {\bf 2}, 3619 (1970).

\bibitem{Keldysh1965} L. V. Keldysh and
Yu. V. Kopaev, Sov. Phys. Solid State {\bf 6}, 2219 (1965).

\bibitem{Andersen2011} O.K. Andersen, and L. Boeri, Annalen der Physik {\bf 1}, 8 (2011).

\bibitem{Liu2010} C. Liu, T. Kondo, R. M. Fernandes, A. D. Palczewski, E. D. Mun, N. Ni, A. N. Thaler, A. Bostwick, E. Rotenberg, J. Schmalian, S. L. Bud'ko, P. C. Canfield and A. Kaminski, Nature Physics {\bf 6}, 419 (2010).

\bibitem{Pratt2011} D.K. Pratt, M.G. Kim, A. Kreyssig, Y.B. Lee, G.S. Tucker, A. Thaler, W. Tian, J.L. Zarestky, S.L. Bud'ko, P.C. Canfield, B.N. Harmon,
A.I. Goldman, and R.J. McQueeney, Phys. Rev. Lett. {\bf 106}, 257001 (2011).

\bibitem{cvetkovic2009} V. Cvetkovic and Z. Tesanovic, Europhys. Lett. {\bf 85}, 37002 (2009).

\bibitem{Chubukov2008} A. V. Chubukov, D. V. Efremov, and I. Eremin, Phys. Rev. B {\bf78} 134512 (2008).

\bibitem{Brydon2009} P. M. R. Brydon and C. Timm, Phys. Rev. B {\bf 80}, 174401 (2009); {\bf 79}, 180504(R) (2009).

\bibitem{wang2009} F. Wang, H. Zhai, Y. Ran, A. Vishwanath, and D.-H. Lee, Phys.
Rev. Lett. {\bf 102}, 047005 (2009).

\bibitem{platt2009} C. Platt, C. Honerkamp, and W. Hanke, New J. Phys. {\bf 11}, 055058
(2009).


\bibitem{Cruz2008} Clarina de la Cruz, Q. Huang, J. W. Lynn, J. Li, W. Ratcliff II, J.L. Zarestky, H.A. Mook, G.F. Chen, J.L. Luo, N.L. Wang, and P. Dai, Nature {\bf 453}, 899 (2008).

\bibitem{Klauss2008} H.-H. Klauss, H. Luetkens, R. Klingeler, C. Hess, F.J. Litterst, M. Kraken, M. M. Korshunov, I. Eremin, S.-L. Drechsler, R. Khasanov, A. Amato, J. Hamann-Borreo, N. Leps, A. Kondrat, G. Behr, J. Werner, B. B\"uchner,
Phys. Rev. Lett. {\bf 101}, 077005 (2008).




\bibitem{Fisher2011} I. R. Fisher, L. Degiorgi, and Z. X. Shen, Rep. Prog. Phys. {\bf 74} 124506 (2011).

\bibitem{Chu2010} J.-H. Chu, J. G. Analytis, K. De Greve, P. L. McMahon,
Z. Islam, Y. Yamamoto, and I. R. Fisher, Science {\bf 329}, 824 (2010).

\bibitem{Tanatar2010} M.A. Tanatar, E.C. Blomberg, A. Kreyssig, M.G. Kim,
N. Ni, A. Thaler, S.L. Bud'ko, P.C. Canfield, A.I.
Goldman, I.I. Mazin, and R. Prozorov, Phys. Rev. B {\bf 81}, 184508 (2010).

\bibitem{Dusza2011} A. Dusza, A. Lucarelli, F. Pfuner, J.-H. Chu, I.R. Fisher,
and L. Degiorgi, EPL 93, 37002 (2011).

\bibitem{Nakajima2011} M. Nakajima, T. Liang, S. Ishida, Y. Tomioka, K. Kihou,
C. H. Lee, A. Iyo, H. Eisaki, T. Kakeshita, T. Ito, and S. Uchida, PNAS {\bf 108}, 12238 (2011).

\bibitem{Matsuda2012} S. Kasahara et al, Nature {\bf 486}, 382 (2012).


\bibitem{Fang2008} C. Fang, H. Yao, W.-F. Tsai, J. P. Hu, and S. A. Kivelson, Phys. Rev. B {\bf 77}, 224509 (2008).

\bibitem{Qi2009} Y. Qi and C. Xu, Phys. Rev. B {\bf 80}, 094402 (2009).

\bibitem{Fernandes2010} R.M. Fernandes, L.H. VanBebber, S. Bhattacharya, P. Chandra, V. Keppens, D. Mandrus, M.A. McGuire, B.C. Sales, A.S. Sefat, and J. Schmalian, Phys. Rev. Lett. {\bf 105}, 157003 (2010).

\bibitem{Yi2011} M. Yi, D. Lu, J.-H Chu, J.G. Analytis, A.P. Sorini,
A.F. Kemper, B. Moritz, S.-K. Mo, R.G. Moore, M.
Hashimoto, W.-S. Lee, Z. Hussain, T.P. Devereaux, I.R. Fisher, and Z.-X. Shen, PNAS {\bf 108}, 6878 (2011).

\bibitem{Kruger2009} F. Kruger, S. Kumar, J. Zaanen, J. van den Brink, Phys. Rev. B {\bf 79}, 054504 (2009).

\bibitem{Lv2010} W. Lv, F. Kr\"uger, and P. Phillips, Phys. Rev. B {\bf 82}, 045125 (2010).

\bibitem{Chen2010} C.-C. Chen, J. Maciejko, A. P. Sorini, B. Moritz, R. R. P. Singh, and T. P. Devereaux, Phys. Rev. B {\bf 82}, 100504 (2010).

\bibitem{Kuroki2009} K.Kuroki, H. Usui1, S. Onari, R. Arita, and H. Aoki, Phys. Rev. B {\bf 79}, 224511 (2009).

\bibitem{Wang2010}  F. Wang, H. Zhai, and D.-H. Lee, Phys. Rev. B {\bf 81}, 184512 (2010).

\bibitem{Maiti2011} S. Maiti, M. M. Korshunov, T. A. Maier, P. J. Hirschfeld, and A. V. Chubukov
Phys. Rev. B {\bf 84}, 224505 (2011).

\bibitem{Thomale2011} R. Thomale, Ch. Platt, W. Hanke, B.A. Bernevig,
Phys. Rev. Lett. {\bf 106}, 187003 (2011).

\bibitem{Kemper2010} A.F. Kemper, T.A. Maier, S. Graser, H.-P. Cheng, P.J. Hirschfeld, and D.J. Scalapino,
New J. Phys. {\bf 12}, 073030 (2010); T.A. Maier, S. Graser, D.J. Scalapino, and P.J. Hirschfeld,
Phys. Rev. B {\bf 79}, 224510 (2009).

\bibitem{Vishwanath2009} Ying Ran, Fa Wang, Hui Zhai, Ashvin Vishwanath, Dung-Hai Lee, Phys. Rev. B {\bf 79}, 014505 (2009).


\bibitem{Yang2010} B.-J. Yang and H.-Y. Kee, Phys. Rev. B {\bf 82}, 195126 (2010).

\bibitem{Morinari2010} T. Morinari, E. Kaneshita, and T. Tohyama, Phys. Rev. Lett. {\bf 105}, 037203 (2010).

\bibitem{Sugimoto2011} K. Sugimoto, E. Kaneshita, T. Tohyama, J. Phys. Soc. Jpn. {\bf 80}, 033706 (2011).

\bibitem{Daghofer2010} M. Daghofer, Q. Luo, R. Yu, D. Yao, A. Moreo, and
E. Dagotto, Phys. Rev. B {\bf 81}, 180514(R) (2010).

\bibitem{Valenzuela2010} B. Valenzuela, E. Bascones, M.J. Calderon, Phys. Rev. Lett. {\bf 105}, 207202 (2010).

\bibitem{Thalmeier2008} Katsunori Kubo, Peter Thalmeier, J. Phys. Soc. Jpn. {\bf 78}, 083704 (2009).


\bibitem{Raghu2008} S. Raghu, Xiao-Liang Qi, Chao-Xing Liu, D. J. Scalapino, and Shou-Cheng Zhang, Phys. Rev. B {\bf 77}, 220503 (2008).

\bibitem{Kontani2012} S. Onari and H. Kontani, Phys. Rev. Lett. {\bf 109}, 137001(2012).

\bibitem{Chuang2010} T.-M. Chuang, M.P. Allan, Jinho Lee, Y. Xie, Ni Ni, S.L. Bud'ko, G.S. Boebinger, P.C. Canfield, and J.C. Davis,
Science {\bf 327}, 181 (2010).

\bibitem{Knolle2010b} J. Knolle, I. Eremin, A. Akbari, and R. Moessner, Phys. Rev. Lett. {\bf 104}, 257001 (2010).

\bibitem{Akbari2010} A. Akbari, J. Knolle, I. Eremin, and R. Moessner, Phys. Rev. B {\bf 82}, 224506 (2010).

\bibitem{Plonka2013} N. Plonka, A.F. Kemper, S. Graser, A.P. Kampf, and T.P. Devereaux, Phys. Rev. B {\bf 88}, 174518 (2013).

\bibitem{Rosenthal13} E. P. Rosenthal, E. F. Andrade, C. J. Arguello, R. M. Fernandes, L. Y. Xing, X. C. Wang, C. Q. Jin, A. J. Millis, and
A. N. Pasupathy, Nature Phys. {\bf 10}, 1038 (2014).

\end{thebibliography}
\end{document}